\documentclass[letterpaper,preprintnumbers,prd,twocolumn,nofootinbib,nobibnotes,showpacs]{revtex4}
\usepackage{graphics,epsfig,subfigure}
\usepackage{color}
\usepackage{amsfonts}
\usepackage{mathrsfs}
\usepackage{epsfig}
\usepackage{graphicx}%
\usepackage{dcolumn}
\usepackage{amsmath}
\usepackage{color}
\usepackage{color}

\begin{document}
\renewcommand{\baselinestretch}{1.3}
\newcommand\beq{\begin{equation}}
\newcommand\eeq{\end{equation}}
\newcommand\beqn{\begin{eqnarray}}
\newcommand\eeqn{\end{eqnarray}}
\newcommand\nn{\nonumber}
\newcommand\fc{\frac}
\newcommand\lt{\left}
\newcommand\rt{\right}
\newcommand\pt{\partial}
\newcommand{\newc}{\newcommand}
\def\eq$#1${\begin{equation}#1\end{equation}}
\def\gat$#1${\begin{gather}#1\end{gather}}
\def\bal$#1${\begin{align}#1\end{align}}
\def\eqarr$#1${\begin{eqnarray}#1\end{eqnarray}}
\newc{\pa}{\partial}
\newc{\alp}{\alpha}
\newc{\gam}{\gamma}
\newc{\Gam}{\Gamma}
\newc{\del}{\delta}
\newc{\eps}{\epsilon}
\newc{\lam}{\lambda}
\newc{\sig}{\sigma}
\newc{\ups}{\upsilon}
\newc{\ome}{\omega}
\newc{\nonum}{\nonumber}
\newc{\vph}{\varphi}
\newc{\tx}{\tilde{x}}
\newc{\paper}[2]{\enquote{\textit{#1}} \cite{#2}}
\newc{\erf}{\text{erf}}

\allowdisplaybreaks

\title{New understanding of Majumdar-Papapetrou and Perj\'es-Israel-Wilson solutions}
\author{Changjun Gao\footnote{gaocj@nao.cas.cn}}

\affiliation{National Astronomical Observatories, Chinese Academy of Sciences, 20A Datun Road, Beijing 100101, China}

\affiliation{School of Astronomy and Space Sciences, University of Chinese Academy of Sciences,
19A Yuquan Road, Beijing 100049, China}

\begin{abstract}
Hartle and Hawking have shown that, except for the case of only point monopoles as sources, both the Majumdar-Papapetrou and the Perj\'es-Israel-Wilson solutions have naked singularities if the space is asymptotically flat. In this article, by using the method of introducing interior solutions, these naked singularities are erased.
\end{abstract}






\maketitle


\section{Introduction}
In 1947, Majumdar \cite{maj:1947} and Papapetrou \cite{pap:1947} studied the source-free Einstein-Maxwell equations in the background of a four dimensional static spacetime. They found that the Einstein-Maxwell equations are eventually reduced to a single Laplace equation in three dimensional Euclidian space. In the weak field limit, this Laplace equation determines exactly the electrostatic potentials. Given the boundary conditions, one can obtain the corresponding MP (Majumdar-Papapetrou) solutions. Among these solutions, the MERN (multiple extreme Reissner-Nordstr$\ddot{\textrm{o}}$m) metric is particular interesting \cite{har:1972,kra:1980}. One of the reasons for this point is that the MERN metric is also the exact solution of
the low energy limit of string theory \cite{ber:1996} which respects supersymmetry \cite{gib:1982,tod:1983}. On the other hand, the MERN solution is one of the few solutions that do not have naked singularities \cite{har:1972} for the source free Einstein-Maxwell equations.

As one of the extensions of MP solutions, the solutions to Einstein-Maxwell
equations together with extremely charged dust source were explored by Das \cite {das:1962}, G$\ddot{\textrm{u}}$rses \cite{gur:1994,gur:1998} and Spruck with Yang \cite{spr:2010}. As another extension of MP solutions, the solutions to Einstein-Maxwell equations with the Einstein cosmological constant, the so-called cosmological multi-black hole solutions, were studied by Kastor et al. \cite{kas:1993,bri:1994}. The cosmological multi-black hole solutions are remarkably interesting because they are time dependent. Therefore, they are used to study the collision of two extreme Reissner-Nordstr$\ddot{\textrm{o}}$m black holes \cite{bri:1994}. Combining the above two extensions, the solutions to Einstein-Maxwell equations with both extremely charged dust and Einstein cosmological constant are constructed by G$\ddot{\textrm{u}}$rses and Himmeto$\bar{\textrm{g}}$lu \cite{gur:2005}. In order to hide the naked singularities, the matter sources in these solutions are placed in multiple shells. Finally, Lemos and Zanchin recognized that the
Majumdar-Papapetrou solutions allow a natural extension to higher dimensional spacetimes \cite{lem:2005a,lem:2005b}. The above mentioned research revolves around the Majumdar-Papapetrou solutions.

In 1971, Perj\'es \cite{per:1971} and in 1972, Israel and Wilson \cite{isr:1972} generalized the techniques of
Majumdar and Papapetrou to find a class of stationary solutions to the
source free Einstein-Maxwell equations. These solutions are named after (PIW) Perj\'es-Israel-Wilson solutions. Same as the case of MP solutions, the Einstein-Maxwell equations are also eventually reduced to a single Laplace equation in three dimensional Euclidian space. But in the weak field limit, this Laplace equation determines exactly not only the electrostatic potentials but also the magnetostatic potentials.
The PIW solutions, as a general class of supersymmetric metrics are discussed by Whitt \cite{whi:1985}. Then Alfredo and Oleg \cite{alf:1998a,alf:1998b} generalized the PIW solutions into the low-energy limit of heterotic string theory.

However, whether for the MP solutions or for the PIW solutions, Hartle and Hawking \cite{har:1972} have shown that, except for the case of only point monopoles as sources, there always exist naked singularities if the space is asymptotically flat. In view of this point, G$\ddot{\textrm{u}}$rses and Himmeto$\bar{\textrm{g}}$lu \cite{gur:2005} placed the matter sources in multiple shells such that the naked singularities are hidden. In this article, with the introduction of the interior solutions, the naked singularities in both MP and PIW solutions are erased because the interior solutions are regular. This constitutes our new understanding of MP and PIW solutions.

The article is organized as follows. In section II, we put forward six MP solutions which are all assigned with exterior and interior solutions. They are for the extreme Reissner-Nordstr$\ddot{\textrm{o}}$m spacetime, the deformed extreme Reissner-Nordstr$\ddot{\textrm{o}}$m spacetime, the spacetime for a dielectric sphere in uniform electric field, the spacetime generated by electric dipole and electric octupole, the general deformed extreme Reissner-Nordstr$\ddot{\textrm{o}}$m spacetime and the spacetime generated by charge distribution on a plane. In section III, we propose three PIW solutions with all of them are assigned with exterior and interior solutions. They are for the spacetime of uniformly magnetized sphere, the spacetime of general electromagnetic object and the spacetime generated by electromagnetic monopoles and dipoles. Finally, section IV gives the conclusion and discussion. Throughout the paper, we adopt the
system of units in which $G=c=\hbar=4\pi\varepsilon_0=1$ and the metric signature $(-, +, +, +)$.

\section{understanding the Majumdar-Papapetrou solutions}
In 1947, Majumdar \cite{maj:1947} and Papapetrou \cite{pap:1947} studied the static spacetime metric with
the conformastatic \cite{syn:1960} form
\begin{eqnarray}\label{MP}
ds^2=-\frac{1}{U\left(\textbf{x}\right)^2}dt^2+U\left(\textbf{x}\right)^2d\mathbf{x}\cdot d\mathbf{x}\;,
\end{eqnarray}
where $\textbf{x}$ denotes the position vector in a flat three dimensional space. In spherical coordinate system, we have
\begin{eqnarray}
d\mathbf{x}\cdot d\mathbf{x}=dr^2+r^2d\theta^2+r^2\sin^2\theta d\phi^2\;.
\end{eqnarray}
They showed that the Einstein-Maxwell equations
\begin{eqnarray}
G_{\mu\nu}&=&-2F_{\mu\lambda}F_{\nu}^{\lambda}+\frac{1}{2}g_{\mu\nu}F_{\alpha\beta}F^{\alpha\beta}\;,
\end{eqnarray}
\begin{eqnarray}
\nabla_{\mu}\left(F^{\mu\nu}\right)=0\;,
\end{eqnarray}
are simplified to be
\begin{eqnarray}\label{lap}
\nabla^2U=0\;, \ \ \ \ \ \ A_{0}=-\frac{1}{U}\;.
\end{eqnarray}
Here $\nabla^2$ is the three dimensional Laplace operator in Euclidian space and $A_0$ is the time-component of electromagnetic four-vector. We note that the expression of $A_0$ is not positive but negative. This is different from the convention of Chandrasekhar \cite{chan:1983}. The difference does not have any effect on the nature of physics because of the gauge invariance of Faraday tensor $F_{\mu\nu}$. If we let
\begin{eqnarray}
U=1+\Phi\;,
\end{eqnarray}
then Eqs.~(\ref{lap}) become
\begin{eqnarray}
\nabla^2\Phi=0\;, \ \ \ \ \ \ A_{0}=-\frac{1}{1+\Phi}\;.
\end{eqnarray}
In the weak field limit $\Phi\ll1$, we have
\begin{eqnarray}
A_{0}=-\frac{1}{1+\Phi}\approx -1+\Phi\sim\Phi\;.
\end{eqnarray}
The minus unit one in the last step is given up because of the gauge invariance of $A_{\mu}$. We know in order to solve the Laplace equation, the boundary condition (BC) must be presumed. Therefore, by appropriately assigning BC, one could get many desired solutions.

The general axisymmetric solution of Laplace equation is
\begin{eqnarray}
\Phi=\sum_{n=0}^{\infty}\left[b_n\left(\frac{r}{a}\right)^{n}+c_{n+1}\left(\frac{a}{r}\right)^{n+1}\right] \cdot P_{n}\left(\cos\theta\right)\;.
\end{eqnarray}
Here $b_n\;, c_{n+1}\;$ are constants and $a$ is the radius of a sphere. $P_{n}\left(\cos\theta\right)$ is the Legendre polynomial of the first kind of order $n$.
Then we obtain some spacetimes as follows.

\subsection{Extreme Reissner-Nordstr$\ddot{\textrm{o}}$m spacetime}
Consider an electric equipotential sphere of radius $a$ that has the electrostatic potential on its surface given by
\begin{eqnarray}
\Phi\left(a\;, \theta\right)=\frac{Q}{a}\;,
\end{eqnarray}
with $Q$ the electric charge inside the sphere. Then the potential at points exterior to the sphere is just the Coulomb potential
\begin{eqnarray}
\Phi_{e}=\frac{Q}{r}\;.
\end{eqnarray}
Inspired by the previous discussion, we propose the spacetime exterior to the sphere is described by
\begin{eqnarray}
U_e=1+\frac{Q}{r}\;,\ \ \ \ A_{0e}=-\frac{1}{U_e}\;,
\end{eqnarray}
which does solve the Einstein-Maxwell equations. In weak field limit, we have $A_{0e}=\Phi_e$. In the interior of the sphere, the spacetime is
\begin{eqnarray}
U_{i}=1+\frac{Q}{a}\;,\ \ \ \ A_{0i}=-\frac{1}{U_i}\;,
\end{eqnarray}
i.e. the Minkowski spacetime. It seems the exterior spacetime is uniquely generated by the electric charge on the sphere. We know it is not the truth. In fact, the Komar mass $M$ of the spacetime is exactly
\begin{equation}\label{km}
 M:=-\frac{1}{8 \pi} \oint_{\mathscr{S}_{t}} \nabla^{\mu} k^{v} \mathrm{~d} S_{\mu v}=Q\;.
\end{equation}
Here the integration is performed on a closed 2-surface $\mathscr{S}_{t}$ in $\Sigma_{t}$. Here $\Sigma_{t}$ provides a $3+1$ foliation $(\Sigma_{t})_{t\in \mathbb{R}}$ of the spacetime. The line element of the surface is
 \begin{equation}\label{2surface}
 \mathrm{d} S_{\mu \nu}=\left(s_{\mu} n_{v}-n_{\mu} s_{v}\right) \sqrt{q} \mathrm{~d}^{2} y\;,
 \end{equation}
 where $n$ is the unit timelike vector normal to $\Sigma_{t}$. $s$ is the unit vector normal to $\mathscr{S}_{t}$ within $\Sigma_{t}$ which is oriented towards the exterior of $\mathscr{S}_{t}$. $y^a=(y^1,y^2)$ are coordinates spanning  $\mathscr{S}_{t}$. $q$ is defined as $q:=det(q_{ab})$ and $q_{ab}$ are the components of the metric induced by $g$ on  $\mathscr{S}_{t}$.  In the process of calculation for the Komar mass, we have closely followed the detailed procedure given by Gourgloulhon \cite{gou:2012}.

 Therefore, the solution is for the extreme Reissner-Nordstr$\ddot{\textrm{o}}$m spacetime. Eq.~(\ref{km}) tells us the mass is positive provided that the charge $Q$ is positive. If the charge is negative, one should propose the spacetime exterior and interior to the sphere as follows
\begin{eqnarray}
U_e=1-\frac{Q}{r}\;,\ \ \ \ U_{i}=1-\frac{Q}{a}\;.
\end{eqnarray}
Then we find $M=-Q$. In other words, the mass is always positive. In all, there are mass and charge distribution on the sphere. The surface density of charge $\sigma_c$ is
\begin{eqnarray}
\sigma_c=-\frac{\partial_r\left(A_{0e}-A_{0i}\right)}{4\pi}|_{r=a}=\frac{Q}{4\pi\left(a+Q\right)^2}\;,
\end{eqnarray}
and the total charge is
\begin{eqnarray}
\int_{0}^{\pi} d\theta \int_{0}^{2\pi} d\phi\sigma_c U_e^2 r^2\sin\theta|_{r=a}=Q\;.
\end{eqnarray}

The surface density of mass $\sigma_m$ is determined by the Darmois-Israel junction conditions \cite{dar:1927,isr:1966,isr:1967,lan:1924}
\begin{eqnarray}
\left[K\right]h_{ij}-\left[K_{ij}\right]=8\pi S_{ij}\;,\ \ \ \ K\equiv K_{i}^{i}\;,
\end{eqnarray}
where the square bracket indicates the jump discontinuity across the sphere, $h_{ij}$ is the induced metric, $S_{ij}$
is the energy momentum tensor localized within the sphere, and all indices pertain to coordinates of the sphere.
Then we find
\begin{eqnarray}
\sigma_m=\frac{Q}{4\pi\left(a+Q\right)^2}=\sigma_c\;.
\end{eqnarray}
Now there is no singularity in this spacetime.

\subsection{Deformed extreme Reissner-Nordstr$\ddot{\textrm{o}}$m spacetime}
Consider a sphere of radius $a$ that has an electrostatic potential on its surface given by
\begin{eqnarray}
\Phi\left(a\;, \theta\right)=\Phi_0\cos^2\theta\;.
\end{eqnarray}
Then the Laplace equation gives the electric potential at points exterior to the sphere
\begin{eqnarray}
\Phi_e=\frac{a\Phi_0}{3r}+\frac{2\Phi_0}{3}\cdot\frac{a^3}{r^3}\cdot P_{2}\left(\cos\theta\right)\;.
\end{eqnarray}
The resulting spacetime exterior to the sphere
\begin{eqnarray}\label{24}
U_e&=&1+\frac{a\Phi_0}{3r}+\frac{2\Phi_0}{3}\cdot\frac{a^3}{r^3}\cdot P_{2}\left(\cos\theta\right)\;,\nonumber\\
A_{0e}&=&-\frac{1}{U_e}\;,
\end{eqnarray}
satisfies the Einstein-Maxwell equations. In the interior of sphere, the spacetime can be taken as
\begin{eqnarray}
U_i&=&1+\frac{\Phi_0}{3}+\frac{2\Phi_0}{3}\cdot\frac{r^2}{a^2}\cdot P_{2}\left(\cos\theta\right)\;,\nonumber\\
A_{0i}&=&-\frac{1}{U_i}\;,
\end{eqnarray}
which also satisfy the Einstein-Maxwell equations. The two metrics match on the surface, $r=a$ continuously. The exterior spacetime is now understood as a deformed extreme Reissner-Nordstr$\ddot{\textrm{o}}$m spacetime. We see the effect of electric quadrupole appears in the exterior spacetime in this case.

The strength of electric field on the surface of the sphere is continuous in the $\theta$ direction while jumping in the radial direction. Therefore, there is charge distribution on the surface of the sphere,
\begin{eqnarray}
\sigma_{c}&=&-\frac{\partial_r\left(A_{0e}-A_{0i}\right)}{4\pi}|_{r=a}\nonumber\\
&=&\frac{3\Phi_0\left[1+10P_2\left(\cos\theta\right)\right]}{4\pi\left[3+\Phi_0+2\Phi_0P_2\left(\cos\theta\right)\right]^2a}\;.
\end{eqnarray}
The total net charge on the sphere is
\begin{eqnarray}
Q_{c}&=&\int_{0}^{\pi} d\theta \int_{0}^{2\pi} d\phi\sigma_c U_e^2 r^2\sin\theta|_{r=a}=\frac{a\Phi_0}{3}\;.
\end{eqnarray}
We find the Komar mass is given by
\begin{eqnarray}
M=Q_c=\frac{a\Phi_0}{3}\;.
\end{eqnarray}
It should be emphasized that, in essence, the Komar mass is contributed by the term of ${a\Phi_0}/{(3r)}$ in the expression of $U_e$. Thus the expression of $U_e$ in Eq.~(\ref{24})
can be rewritten as
\begin{eqnarray}
U_e&=&1+\frac{M}{r}+\frac{2Q_c a^2}{r^3}\cdot P_{2}\left(\cos\theta\right)\;.
\end{eqnarray}
It tells us the anisotropy of the spacetime is caused by not the mass, but the electric charge. In other words, the mass is uniformly distributed on the sphere. However, it is not the case for the electric charge. Now there is no singularity in this spacetime.
\subsection{Dielectric sphere in uniform electric field}
Consider a dielectric sphere of radius $a$ with the dielectric constant $\epsilon$ which is placed in a uniform electric field of strength $E_0$.
The electric potential is
\begin{eqnarray}
\Phi_i=-\frac{3}{2+\epsilon}\,E_0\,r\,\cos\theta\;,
\end{eqnarray}
for $r\leq a$ and
\begin{eqnarray}
\Phi_e=-E_0\, r\,\cos\theta + \left(\frac{\epsilon-1}{\epsilon+2}\right)\cdot\,E_0\,\frac{a^{\,3}}{r^{\,2}}\cos\theta\;,
\end{eqnarray}
for $r>a$.

The resulting spacetime is given by
\begin{eqnarray}
U_i=1-\frac{3}{2+\epsilon}\,E_0\,r\,\cos\theta\;,\ \ \ \ A_{0i}=-\frac{1}{U_i}\;,
\end{eqnarray}
for $r\leq a$ and
\begin{eqnarray}
U_e&=&1-E_0r\cos\theta+\frac{\epsilon-1}{\epsilon+2}\cdot \frac{E_0a^{3}}{r^{2}}\cos\theta\;,\nonumber\\ A_{0e}&=&-\frac{1}{U_e}\;,
\end{eqnarray}
for $r>a$. The two spacetimes satisfy the Einstein-Maxwell equations. The second and the third term in the expression of $U$ stem from the uniform electric field and the electric diploe, respectively.

The strength of electric field on the surface of the sphere is continuous in the $\theta$ direction while jumping in the radial direction. Therefore, there is charge distribution on the surface of the sphere,
\begin{eqnarray}
\sigma_{c}&=&-\frac{\partial_r\left(A_{0e}-A_{0i}\right)}{4\pi}|_{r=a}\nonumber\\
&=&\frac{3\left(\epsilon-1\right)\left(\epsilon+2\right)E_0\cos\theta}{4\pi\left(2+\epsilon-3E_0a\cos\theta\right)^2}\;.
\end{eqnarray}
The total net charge on the sphere is
\begin{eqnarray}
Q_{c}&=&\int_{0}^{\pi} d\theta \int_{0}^{2\pi} d\phi\sigma_c U_e^2 r^2\sin\theta|_{r=a}=0\;.
\end{eqnarray}
Let $E_0=0$, we find the Komar mass is also vanishing. Namely, there is no mass distribution on the sphere. Thus this spacetime is generated uniquely by the electric charges. These kind of results are not puzzling. As another example, if we let the mass term vanishes in the Reissner-Nordstr$\ddot{\textrm{o}}$m spacetime, we are left with a spacetime generated uniquely by the electric charge. In the next subsection, we shall find this situation again.

\subsection{Spacetime generated by electric dipole and electric octupole}
Consider a sphere of radius $a$ that has an electrostatic potential on its surface given by
\begin{eqnarray}
\Phi\left(a\;, \theta\right)=\Phi_0\cos\left(3\theta\right)\;.
\end{eqnarray}
The electric potential at points exterior to the sphere is
\begin{eqnarray}
\Phi_e=-\frac{3\Phi_0a^2}{5r^2}P_1\left(\cos\theta\right)+\frac{8\Phi_0}{5}\cdot\frac{a^4}{r^4}\cdot P_{3}\left(\cos\theta\right)\;,
\end{eqnarray}
and at points interior to the sphere is
\begin{eqnarray}
\Phi_i=-\frac{3\Phi_0 r}{5a}P_1\left(\cos\theta\right)+\frac{8\Phi_0}{5}\cdot\frac{r^3}{a^3}\cdot P_{3}\left(\cos\theta\right)\;,
\end{eqnarray}
The resulting spacetime exterior to the sphere
\begin{eqnarray}
U_e&=&1+\Phi_e\;,\nonumber\\
A_{0e}&=&-\frac{1}{U_e}\;,
\end{eqnarray}
satisfies the Einstein-Maxwell equations. In the interior of sphere, the spacetime can be taken as
\begin{eqnarray}
U_i&=&1+\Phi_i\;,\nonumber\\
A_{0i}&=&-\frac{1}{U_i}\;,
\end{eqnarray}
which also satisfy the Einstein-Maxwell equations. The two metrics match on the surface, $r=a$. The exterior spacetime is contributed by an electric dipole and electric octupole. The strength of electric field on the surface of the sphere is continuous in the $\theta$ direction while jumping in the radial direction. Therefore, there is charge distribution on the surface of the sphere,

\begin{eqnarray}
\sigma_{c}&=&-\frac{\partial_r\left(A_{0e}-A_{0i}\right)}{4\pi}|_{r=a}\nonumber\\
&=&\frac{\Phi_0\cos\theta\left(-93+140\cos^2\theta\right)}{20\pi a\left(1-3\Phi_0\cos\theta+4\Phi_0\cos^3\theta\right)^2}\;.
\end{eqnarray}
The total net charge on the sphere is
\begin{eqnarray}
Q_{c}&=&\int_{0}^{\pi} d\theta \int_{0}^{2\pi} d\phi\sigma_c U_e^2 r^2\sin\theta|_{r=a}=0\;.
\end{eqnarray}
We find the Komar mass is also vanishing. Namely, there is no mass distribution on the sphere.
\subsection{General deformed extreme Reissner-Nordstr$\ddot{\textrm{o}}$m spacetime}

In view of above discussions, we assume the deformed extreme Reissner-Nordstr$\ddot{\textrm{o}}$m black hole immersed in static electric field is described by
\begin{eqnarray}
U_e&=&\cdot\cdot\cdot+b_2\cdot\frac{r^2}{a^2}\cdot P_2\left(\cos\theta\right)+b_1\cdot\frac{r}{a}\cdot P_1\left(\cos\theta\right)\nonumber\\&&+1+c_1\cdot\frac{a}{r}\cdot P_0+c_2\cdot\frac{a^{2}}{r^{2}}\cdot P_1\left(\cos\theta\right)\nonumber\\&&+c_3\cdot\frac{a^{3}}{r^{3}}\cdot P_2\left(\cos\theta\right)+\cdot\cdot\cdot\;,\nonumber\\
A_{0e}&=&-\frac{1}{U_e}\;,
\end{eqnarray}
when $r>a$. It includes the contributions from uniform electric field ($b_1$ term), the electric charge ($c_1$ term), the electric dipole ($c_2$ term) and the electric quadrupole ($c_3$ term) etc.

When $r<a$, the spacetime can be taken as
\begin{eqnarray}
U_i&=&1+c_1+\left(b_1+c_2\right)\cdot\frac{r}{a}\cdot P_1\left(\cos\theta\right)\nonumber\\&&+\left(b_2+c_3\right)\cdot\frac{r^2}{a^2}\cdot P_2\left(\cos\theta\right)+\cdot\cdot\cdot\;,\nonumber\\
A_{0i}&=&-\frac{1}{U_i}\;.
\end{eqnarray}
The two metrics match on the surface of sphere continuously. The surface density of charge on the sphere is
\begin{eqnarray}
\sigma_{c}&=&-\frac{\partial_r\left(A_{0e}-A_{0i}\right)}{4\pi}|_{r=a}\nonumber\\
&=&\frac{\left[c_1+3c_2P_1\left(\theta\right)+5c_3P_2\left(\theta\right)+\cdot\cdot\cdot\right]}{4\pi a U_e^2|_{r=a}}\;.
\end{eqnarray}
Then the total net charge on the sphere is
\begin{eqnarray}
Q_{c}&=&\int_{0}^{\pi} d\theta \int_{0}^{2\pi} d\phi\sigma_c U_e^2 r^2\sin\theta|_{r=a}=c_1 a\;.
\end{eqnarray}
The Komar mass is exactly equal to the electric charge and it is uniformly distributed on the sphere.

\subsection{Spacetime generated by charge distribution on a plane}
In above subsections, we solve the Laplace equation with boundary conditions on a sphere. In this subsection, we consider the Laplace equation with boundary conditions on an infinite plane. To this end, we work in the rectangular Cartesian coordinates by using the method of separation for variables.

Our solution is given by
\begin{eqnarray}\label{24}
U_{e}&=&1-b_0^2\sin\omega_1x\sin\omega_2ye^{-\sqrt{\omega_1^2+\omega_2^2}z}\;,\nonumber\\
A_{0e}&=&-\frac{1}{U_{e}}\;,
\end{eqnarray}
for $z\geq 0$ and

\begin{eqnarray}\label{24}
U_{i}&=&1-b_0^2\sin\omega_1x\sin\omega_2ye^{\sqrt{\omega_1^2+\omega_2^2}z}\;,\nonumber\\
A_{0i}&=&-\frac{1}{U_{i}}\;.
\end{eqnarray}
for $z<0$. Here $\omega_1$, $\omega_2$ and $b_0$ are all positive constants. In order that the surface density of electric charge on the $z=0$ plane is finite, we require $b_0<1$. The two spacetimes match on the plane of $z=0$ continuously and they are regular everywhere.
The strength of electric field on the plane is continuous in the $x$ and $y$ directions while jumping in the $z$ direction. Therefore, there is charge distribution on the $z=0$ plane,
\begin{eqnarray}
\sigma_{c}&=&-\frac{\partial_z\left(A_{0e}-A_{0i}\right)}{4\pi}|_{z=0}\nonumber\\
&=&-\frac{b_0^2\sin\omega_1x\sin\omega_2y\sqrt{\omega_1^2+\omega_2^2}}{2\pi\left(1-b_0^2\sin\omega_1x\sin\omega_2y\right)^2}\;.
\end{eqnarray}
There is no mass distribution on the plane.

\section{understanding the Perj\'es-Israel-Wilson solutions}
In 1971, Perj\'es \cite{per:1971} and in 1972, Israel and Wilson \cite{isr:1972}, independently, generalized the methods of
Majumdar and Papapetrou and found a class of stationary solutions
to the source free Einstein-Maxwell equations. Their solutions have
the metric
\begin{eqnarray}\label{PIW}
ds^2=-\frac{1}{|U|^2}\left(dt+\mathbf{\Omega}\cdot d\bf{x}\right)^2+|U|^2d\mathbf{x}\cdot d\mathbf{x}\;,
\end{eqnarray}
where $U$ becomes now a complex function and satisfies the Laplace equation in three dimensional Euclidian space
 \begin{eqnarray}
\nabla^2U=0\;.
\end{eqnarray}
The three dimensional vector $\mathbf{\Omega}$ is determined by
\begin{eqnarray}\label{omega}
\nabla\times \mathbf{\Omega}=i\left(U\nabla \bar{U}-\bar{U}\nabla U\right)\;.
\end{eqnarray}
The electrostatic potential $\Phi$ and magnetic scalar potential $\chi$ are defined by the relations
\begin{eqnarray}
F_{ti}=\partial_i\Phi\;,\ \ \ \ \ F^{ij}=|U|^2\varepsilon^{ijk}\partial_k\chi\;,
\end{eqnarray}
and they are related to the function $U$ as follows
\begin{eqnarray}
\Phi+i\chi=-\frac{1}{U}\;.
\end{eqnarray}
It is apparent the Perj\'es-Israel-Wilson solutions would reduce to the Majumdar-Papapetrou solutions when the magnetic scalar potential vanishes. In the next subsections, we consider three PIW spacetimes with both exterior and interior spacetimes.
 \subsection{Spacetime of uniformly magnetized sphere}
Consider the scalar magnetic potential produced by a uniformly magnetized sphere. The scalar magnetic potential
is given by
\begin{eqnarray}
\chi_{iw}=\frac{M_0a^2}{3}\frac{r}{a^2}\cos\theta\;,
\end{eqnarray}
for $r\leq a$ and
\begin{eqnarray}
\chi_{ew}=\frac{M_0a^2}{3}\frac{a}{r^2}\cos\theta\;,
\end{eqnarray}
for $r>a$. $M_0$ is the uniform permanent magnetization. Motivated by above potentials, we propose the spacetime generated by a uniformly magnetized sphere is given by
\begin{eqnarray}
U_i=-i\left(1+\frac{M_0a^2}{3}\frac{r}{a^2}\cos\theta\right)\;,\ \chi_i=\frac{i}{U_i}\;,\ \mathbf{\Omega_i}=0\;,
\end{eqnarray}
for $r\leq a$ and
\begin{eqnarray}
U_e=-i\left(1+\frac{M_0a^2}{3}\frac{a}{r^2}\cos\theta\right)\;,\ \chi_e=\frac{i}{U_e}\;,\ \mathbf{\Omega_e}=0\;,
\end{eqnarray}
for $r>a$. The spacetime satisfies the Einstein-Maxwell equations. In the weak field limit, we have $\chi_{(i,e)}=\chi_{(i,e)w}$.
The strength of magnetic field on the surface of the sphere is continuous in the $\theta$ direction while jumping in the radial direction. Therefore, there is magnetic charge distribution on the surface of the sphere,
\begin{eqnarray}
\sigma_{c}&=&-\frac{\partial_r\left(\chi_{e}-\chi_{i}\right)}{4\pi}|_{r=a}\nonumber\\
&=&\frac{9M_0\cos\theta}{4\pi\left(3+M_0a\cos\theta\right)^2}\;.
\end{eqnarray}
Then the total net magnetic charge on the sphere is
\begin{eqnarray}
Q_{c}&=&\int_{0}^{\pi} d\theta \int_{0}^{2\pi} d\phi\sigma_c U_e^2 r^2\sin\theta|_{r=a}=0\;.
\end{eqnarray}
The Komar mass is also zero.

\subsection{General electromagnetic object}

We suppose a spacetime exterior to a sphere is described by
\begin{eqnarray}
U_e&=&\alpha_e-i\beta_e\;,
\end{eqnarray}
with
\begin{eqnarray}
\alpha_e &=&b_0+\sum_{n=0}^{\infty}c_{n+1}\left(\frac{a}{r}\right)^{n+1}\cdot P_{n}\left(\cos\theta\right)\;,
\end{eqnarray}
\begin{eqnarray}
\beta_e&=&p_0+\sum_{n=0}^{\infty}q_{n+1}\left(\frac{a}{r}\right)^{n+1}\cdot P_{n}\left(\cos\theta\right)\;.
\end{eqnarray}
The solution of Eq.~(\ref{omega}) for $\mathbf{\Omega}$ can be taken as
\begin{eqnarray}
\mathbf{\Omega_e}&=&\sum_{n=1}^{\infty}\mathfrak{O}_{en}\mathbf{\textbf{e}}_{\phi}\;,
\end{eqnarray}
with
\begin{eqnarray}
\mathfrak{O}_{e1}&=&\frac{2a}{r\sin\theta}\left[\cos\theta\left(p_0c_1-b_0q_1\right)+w\right]\;,\nonumber\\
\mathfrak{O}_{e2}&=&\frac{2a^2\sin\theta}{r^2}\left(b_0q_2-p_0c_2\right)\;,\nonumber\\
\mathfrak{O}_{e3}&=&-\frac{a^3\sin\theta}{r^3}\left[-3\cos\theta \left(b_0q_3-p_0c_3\right)+c_2q_1-q_2c_1\right]\;,\nonumber\\
\mathfrak{O}_{e4}&=&-\frac{a^4\sin\theta}{r^4}\left[-5\cos^2\theta \left(b_0q_4-p_0c_4\right)\right.\nonumber\\&&\left.+2\cos\theta \left(c_3q_1-q_3c_1\right)+b_0q_4-p_0c_4\right]\;,\nonumber\\
\mathfrak{O}_{e5}&=&-\frac{a^5\sin\theta}{4r^5}\left[-35\cos^3\theta \left(b_0q_5-p_0c_5\right)\right.\nonumber\\&&\left.+3\cos^2\theta \left(5q_1c_4-5c_1q_4+c_3q_2-q_3c_2\right)\right.\nonumber\\&&\left.+15\cos\theta\left(q_5b_0-c_5p_0\right)\right.\nonumber\\&&\left.
+c_3q_2-q_3c_2+3c_1q_4-3q_1c_4\right]\;,\nonumber\\
&&\cdot\cdot\cdot\cdot\cdot\cdot\;.
\end{eqnarray}

In the interior of the sphere, $U$ is taken as

\begin{eqnarray}
U_i&=&\alpha_i-i\beta_i\;,
\end{eqnarray}
with
\begin{eqnarray}
\alpha_i&=&b_0+\sum_{n=0}^{\infty}c_{n+1}\left(\frac{r}{a}\right)^{n}\cdot P_{n}\left(\cos\theta\right)\;,
\end{eqnarray}
\begin{eqnarray}
\beta_i&=&p_0+\sum_{n=0}^{\infty}q_{n+1}\left(\frac{r}{a}\right)^{n}\cdot P_{n}\left(\cos\theta\right)\;.
\end{eqnarray}
The solution of Eq.~(\ref{omega}) for $\mathbf{\Omega}$ can be taken as
\begin{eqnarray}
\mathbf{\Omega_i}&=&\sum_{n=1}^{\infty}\mathfrak{O}_{in}\mathbf{\textbf{e}}_{\phi}\;,
\end{eqnarray}
with
\begin{eqnarray}
\mathfrak{O}_{i1}&=&\frac{as}{r\sin\theta}\;,\nonumber\\
\mathfrak{O}_{i2}&=&-\frac{r\sin\theta\left(q_2b_0
+q_2c_1-c_2p_0-c_2q_1\right)}{a}\;,\nonumber\\
\mathfrak{O}_{i3}&=&-\frac{2r^2\sin\theta\cos\theta\left(q_3c_1
+q_3b_0-c_3q_1-c_3p_0\right)}{a^2}\;,\nonumber\\
\mathfrak{O}_{i4}&=&-\frac{r^3\sin\theta}{4a^3}\left[3q_4c_1-3c_4q_1+3q_4b_0+c_3q_2\right.\nonumber\\&&\left.-c_2q_3-3c_4p_0
+3\cos^2\theta\cdot\left(c_2q_3-c_3q_2\right.\right.\nonumber\\&&\left.\left.-5c_4q_1-5c_4p_0+5q_4c_1+5q_4b_0\right)\right]\;,\nonumber\\
&&\cdot\cdot\cdot\cdot\cdot\cdot\;,
\end{eqnarray}
with $s$ an integration constant. The interior and exterior metrics should match on the sphere. The boundary conditions are
\begin{eqnarray}
{U_i}|_{r=a}={U_e}|_{r=a}\;,\ \ \ \ \mathbf{\Omega_i}|_{r=a}=\mathbf{\Omega_e}|_{r=a}\;.
\end{eqnarray}
It is very obvious that the first condition is satisfied. But it is very hard to detail the second boundary condition. So in the next section, we shall focus on a very simple case, namely, the spacetime generated by electromagnetic monopoles and dipoles.

 \subsection{Spacetime generated by the electric monopole, electric dipole and their magnetic counterparts}

We suppose a spacetime exterior to a sphere is generated by the electric monopole, electric dipole and their magnetic counterparts as follows
\begin{eqnarray}
U_e&=&\alpha_e-i\beta_e\;,
\end{eqnarray}
with
\begin{eqnarray}
\alpha_e &=&b_0+c_1\cdot\frac{a}{r}+c_2\cdot\frac{a^{2}}{r^{2}}\cos\theta\;,
\end{eqnarray}
\begin{eqnarray}
\beta_e&=&p_0+q_1\cdot\frac{a}{r}+q_2\cdot\frac{a^{2}}{r^{2}}\cos\theta\;.
\end{eqnarray}
The solution of Eq.~(\ref{omega}) for $\mathbf{\Omega}$ can be taken as
\begin{eqnarray}
\mathbf{\Omega_e}&=&-\frac{2a\left[\cos\theta\left(p_0c_1-b_0q_1\right)+w\right]}{r\sin\theta}\mathbf{\textbf{e}}_{\phi}\nonumber\\&&+\frac{2a^2\sin\theta\left(c_2p_0
-q_2b_0\right)}{r^2}\mathbf{\textbf{e}}_{\phi}\nonumber\\&&
+\frac{a^3\sin\theta\left(c_2q_1-q_2c_1\right)}{r^3}\mathbf{\textbf{e}}_{\phi}\;.
\end{eqnarray}
The metric is
\begin{eqnarray}
ds^2&=&-\left(\alpha_e^2+\beta_e^2\right)^{-1}\left(dt+\mathbf{\Omega_e}\cdot d\bf{x}\right)^2\nonumber\\&&+\left(\alpha_e^2+\beta_e^2\right)d\mathbf{x}\cdot d\mathbf{x}\;,
\end{eqnarray}
where
\begin{eqnarray}
&&{\mathbf{\Omega_e}}\cdot d \mathbf{x}=d\phi\left\{-2a\left[\cos\theta\left(p_0c_1-b_0q_1\right)+w\right]\right.\nonumber\\&&\left.+\frac{2a^2\sin^2\theta\left(c_2p_0
-q_2b_0\right)}{r}+\frac{a^3\sin^2\theta\left(c_2q_1-q_2c_1\right)}{r^2}\right\}\;.
\end{eqnarray}
Here, $w$ is a dimensionless integration constant. The electrostatic potential $\Phi$ and scalar magnetic potential $\chi$ are
\begin{eqnarray}
\Phi_e=\frac{-\alpha_e}{\alpha_e^2+\beta_e^2}\;,\ \ \ \ \chi_e=\frac{-\beta_e}{\alpha_e^2+\beta_e^2}\;.
\end{eqnarray}

In the interior of the sphere, $U$ is taken as

\begin{eqnarray}
U_i&=&\alpha_i-i\beta_i\;,
\end{eqnarray}
with
\begin{eqnarray}
\alpha_i&=&b_0+c_1+c_2\cdot\frac{r}{a}\cos\theta\;,
\end{eqnarray}
\begin{eqnarray}
\beta_i&=&p_0+q_1+q_2\cdot\frac{r}{a}\cos\theta\;.
\end{eqnarray}
The solution of Eq.~(\ref{omega}) for $\mathbf{\Omega}$ can be taken as
\begin{eqnarray}
\mathbf{\Omega_i}&=&-\frac{r\sin\theta\left(q_2b_0
+q_2c_1-c_2p_0-c_2q_1\right)}{a}\mathbf{\textbf{e}}_{\phi}\nonumber\\&&-\frac{as}{r\sin\theta}\mathbf{\textbf{e}}_{\phi}\;,
\end{eqnarray}
with $s$ an integration constant. The electrostatic potential $\Phi$ and scalar magnetic potential $\chi$ are
\begin{eqnarray}
\Phi_i=\frac{-\alpha_i}{\alpha_i^2+\beta_i^2}\;,\ \ \ \ \chi_i=\frac{-\beta_i}{\alpha_i^2+\beta_i^2}\;.
\end{eqnarray}
The interior and exterior metrics should match on the surface of sphere. Then
\begin{eqnarray}
U_i|_{r=a}={U_e}|_{r=a}\;,\ \ \ \ \mathbf{\Omega_i}|_{r=a}=\mathbf{\Omega_e}|_{r=a}\;,
\end{eqnarray}
gives
\begin{eqnarray}
s=2w\;,\ \ \ p_0=-\frac{2q_1}{3}\;,\ \ \ b_0=-\frac{2c_1}{3}\;.
\end{eqnarray}
The surface density on the sphere is
\begin{eqnarray}
\sigma_{ec}&=&-\frac{\partial_r\left(\Phi_{e}-\Phi_{i}\right)}{4\pi}|_{r=a}\nonumber\\
&=&\frac{9\left(c_1+3c_2\cos\theta\right)}{4\pi a\left[\left(c_1+3c_2\cos\theta\right)^2+\left(q_1+3q_2\cos\theta\right)^2\right]}\;,
\end{eqnarray}
for the electric charge and
\begin{eqnarray}
\sigma_{mc}&=&-\frac{\partial_r\left(\chi_{e}-\chi_{i}\right)}{4\pi}|_{r=a}\nonumber\\
&=&\frac{9\left(q_1+3q_2\cos\theta\right)}{4\pi a\left[\left(c_1+3c_2\cos\theta\right)^2+\left(q_1+3q_2\cos\theta\right)^2\right]}\;,
\end{eqnarray}
for the magnetic charge. Then the net electric charge and net magnetic charge on the sphere are
\begin{eqnarray}
&&Q_{ec}=\int_{0}^{\pi} d\theta \int_{0}^{2\pi} d\phi\sigma_{ec} |U_e|^2 r^2\sin\theta|_{r=a}=c_1a\;,\\
&&Q_{mc}=\int_{0}^{\pi} d\theta \int_{0}^{2\pi} d\phi\sigma_{mc} |U_e|^2 r^2\sin\theta|_{r=a}=q_1a\;.
\end{eqnarray}

In the weak field limit we have the weak electrostatic potential $\Phi_{ew}$ and weak magnetic potential $\chi_{wm}$ as follows
\begin{eqnarray}
\Phi_{we}&=&\frac{Q_{ec}}{r}+\frac{c_2a^{2}}{r^{2}}\cos\theta\;,\nonumber\\ \chi_{wm}&=&\frac{Q_{mc}}{r}+\frac{q_2a^{2}}{r^{2}}\cos\theta\;.
\end{eqnarray}
These potentials are exactly produced by the electric monopole, electric dipole and their magnetic counterparts, respectively. Thus the magnitude of electric dipole $D_{ed}$ and magnetic dipole $D_{md}$ are
\begin{eqnarray}
D_{ed}=c_2a^2\;,\ \ \ \ \ {D_{md}}=q_2a^2\;.
\end{eqnarray}
We require the spacetime is asymptotically Minkowski in spatial infinity. Then we find the integration constant $w$ should vanishes,
\begin{eqnarray}
w=0\;.
\end{eqnarray}
We eventually obtain the metric components for the exterior spacetime
\begin{eqnarray}
|U|_e^2&=&\left(-\frac{2Q_{ec}}{3a}+\frac{Q_{ec}}{r}+\frac{D_{ed}\cos\theta}{r^2}\right)^2\nonumber\\&&
+\left(-\frac{2Q_{mc}}{3a}+\frac{{Q_{mc}}}{r}+\frac{{D_{md}}\cos\theta}{r^2}\right)^2\;,
\end{eqnarray}
\begin{eqnarray}\label{93a}
&&{\mathbf{\Omega_e}}\cdot d \mathbf{x}=\left[\frac{\left(D_{ed}{Q_{mc}}-D_{md}Q_{ec}\right)\sin^2\theta}{r^2}\right.\nonumber\\&&\left.
-\frac{4\left(D_{ed}{Q_{mc}}-{D_{md}}Q_{ec}\right)\sin^2\theta}{3ar}\right]d\phi\;,
\end{eqnarray}
and the metric components for the interior spacetime
\begin{eqnarray}
|U|_{i}^2&=&\left(\frac{Q_{ec}}{3a}+\frac{D_{ed}r\cos\theta}{a^3}\right)^2\nonumber\\&&
+\left(\frac{{Q_{mc}}}{3a}+\frac{{D_{md}r}\cos\theta}{a^3}\right)^2\;,
\end{eqnarray}
\begin{eqnarray}
{\mathbf{\Omega_i}}\cdot d \mathbf{x}&=&{\frac{r^2}{3a^4}\sin^2\theta\left(D_{md}Q_{ec}-D_{ed}{Q_{mc}}\right)}d\phi\;.
\end{eqnarray}
In the end, we examine whether the closed timelike curves (CTCs) and closed null curves (CNCs) exist in the exterior spacetime. These curves would arise when the metric component $g_{\phi\phi}$ occurs to be negative and vanishing, respectively. It is necessary to consider first of all, the regions near the axis of
symmetry, i.e. for $\theta\to 0$ or for $\theta\to
\pi$ for the values of $r>a$. Eq.~(\ref{93a}) tells us when $\theta\to 0$ or for $\theta\to
\pi$, the $dtd\varphi$ term disappears and $g_{\varphi\varphi}$ is always positive. Therefore, there is no closed timelike curves near the axis of symmetry. So we shall focus on the case of $\theta=\pi/2$ in the next. Then $g_{\phi\phi}=0$ gives two roots for electric dipole $D_{ed}$ denoted by $D_{1ed}$ and $D_{2ed}$,

\begin{eqnarray}
D_{1ed}&=&\frac{1}{3aQ_{mc}\left(3a-4r\right)}\left[4\left(Q_{ec}^2+Q_{mc}^2\right)r^3\right.\nonumber\\&&\left.+3a\left(3aQ_{ec}^2+3aQ_{mc}^2-4Q_{ec}D_{md}\right)r
\right.\nonumber\\&&\left.+9a^2Q_{ec}D_{md}-12a\left(Q_{ec}^2+Q_{mc}^2\right)r^2\right]\;,
\end{eqnarray}
\begin{eqnarray}
D_{2ed}&=&D_{1ed}+\frac{2r\left(Q_{ec}^2+Q_{mc}^2\right)\left(2r-3a\right)^2}{3aQ_{mc}\left(4r-3a\right)}\;.
\end{eqnarray}
We find when the electric dipole $D_{ed}$ satisfies
\begin{eqnarray}\label{cd}
D_{1ed}<D_{ed}<D_{2ed}\;,
\end{eqnarray}
we always have
\begin{eqnarray}
g_{\phi\phi}>0\;,
\end{eqnarray}
provided that $Q_{mc}>0$ and $r\neq 3a/2$. In other words, there would be no CTCs and CNCs in the equatorial plane when the condition of Eq.~(\ref{cd}) is satisfied.

\section{conclusion and discussion}
In all, both the MP spacetimes and the PIW spacetimes are the solutions of source-free Einstein-Maxwell equations. In essence, the MP spacetimes are generated by electrostatic potentials and the PIW solutions are generated by both the electrostatic and magnetostatic potentials. In any case, the Einstein-Maxwell equations are reduced to a single Laplace equation in three dimensional Euclidian space. To give a physical and specific solution, the boundary conditions must be presumed.

In this article, by introducing the interior solutions within a sphere of radius $a$, we construct several exact solutions for the MP and PIW spacetimes. All these spacetimes are regular everywhere and some of them are also asymptotically flat. Furthermore, there is no singularities in these spacetimes. The exterior and interior spacetimes match on the surface of the sphere continuously. So there is charge distribution on the surface. The spacetimes are generated by electric and magnetic monopoles, dipoles, quadrupoles, static uniform fields, static non-uniform fields and so on. Therefore, they are the counterparts of the electromagnetic objets for classical electrodynamics in General Relativity. Thus it is interesting to consider the application of these solutions in the high-energy astrophysics, such as neutron star physics, in the future.

\section*{ACKNOWLEDGMENTS}

I am grateful to Prof. Luca Bombelli for pointing out the typos in names and references. This work is partially supported by the Strategic Priority Research Program ``Multi-wavelength Gravitational Wave Universe'' of the
CAS, Grant No. XDB23040100 and the NSFC under grants 11633004, 11773031.

\newcommand\arctanh[3]{~arctanh.{\bf ~#1}, #2~ (#3)}
\newcommand\ARNPS[3]{~Ann. Rev. Nucl. Part. Sci.{\bf ~#1}, #2~ (#3)}
\newcommand\AL[3]{~Astron. Lett.{\bf ~#1}, #2~ (#3)}
\newcommand\AP[3]{~Astropart. Phys.{\bf ~#1}, #2~ (#3)}
\newcommand\AJ[3]{~Astron. J.{\bf ~#1}, #2~(#3)}
\newcommand\GC[3]{~Grav. Cosmol.{\bf ~#1}, #2~(#3)}
\newcommand\APJ[3]{~Astrophys. J.{\bf ~#1}, #2~ (#3)}
\newcommand\APJL[3]{~Astrophys. J. Lett. {\bf ~#1}, L#2~(#3)}
\newcommand\APJS[3]{~Astrophys. J. Suppl. Ser.{\bf ~#1}, #2~(#3)}
\newcommand\JHEP[3]{~JHEP.{\bf ~#1}, #2~(#3)}
\newcommand\JMP[3]{~J. Math. Phys. {\bf ~#1}, #2~(#3)}
\newcommand\JCAP[3]{~JCAP {\bf ~#1}, #2~ (#3)}
\newcommand\LRR[3]{~Living Rev. Relativity. {\bf ~#1}, #2~ (#3)}
\newcommand\MNRAS[3]{~Mon. Not. R. Astron. Soc.{\bf ~#1}, #2~(#3)}
\newcommand\MNRASL[3]{~Mon. Not. R. Astron. Soc.{\bf ~#1}, L#2~(#3)}
\newcommand\NPB[3]{~Nucl. Phys. B{\bf ~#1}, #2~(#3)}
\newcommand\CMP[3]{~Comm. Math. Phys.{\bf ~#1}, #2~(#3)}
\newcommand\CQG[3]{~Class. Quantum Grav.{\bf ~#1}, #2~(#3)}
\newcommand\PLB[3]{~Phys. Lett. B{\bf ~#1}, #2~(#3)}
\newcommand\PRL[3]{~Phys. Rev. Lett.{\bf ~#1}, #2~(#3)}
\newcommand\PR[3]{~Phys. Rep.{\bf ~#1}, #2~(#3)}
\newcommand\PRd[3]{~Phys. Rev.{\bf ~#1}, #2~(#3)}
\newcommand\PRD[3]{~Phys. Rev. D{\bf ~#1}, #2~(#3)}
\newcommand\RMP[3]{~Rev. Mod. Phys.{\bf ~#1}, #2~(#3)}
\newcommand\SJNP[3]{~Sov. J. Nucl. Phys.{\bf ~#1}, #2~(#3)}
\newcommand\ZPC[3]{~Z. Phys. C{\bf ~#1}, #2~(#3)}
\newcommand\IJGMP[3]{~Int. J. Geom. Meth. Mod. Phys.{\bf ~#1}, #2~(#3)}
\newcommand\IJMPD[3]{~Int. J. Mod. Phys. D{\bf ~#1}, #2~(#3)}
\newcommand\IJMPA[3]{~Int. J. Mod. Phys. A{\bf ~#1}, #2~(#3)}
\newcommand\GRG[3]{~Gen. Rel. Grav.{\bf ~#1}, #2~(#3)}
\newcommand\EPJC[3]{~Eur. Phys. J. C{\bf ~#1}, #2~(#3)}
\newcommand\PRSL[3]{~Proc. Roy. Soc. Lond.{\bf ~#1}, #2~(#3)}
\newcommand\AHEP[3]{~Adv. High Energy Phys.{\bf ~#1}, #2~(#3)}
\newcommand\Pramana[3]{~Pramana.{\bf ~#1}, #2~(#3)}
\newcommand\PTP[3]{~Prog. Theor. Phys{\bf ~#1}, #2~(#3)}
\newcommand\APPS[3]{~Acta Phys. Polon. Supp.{\bf ~#1}, #2~(#3)}
\newcommand\ANP[3]{~Annals Phys.{\bf ~#1}, #2~(#3)}
\newcommand\RPP[3]{~Rept. Prog. Phys. {\bf ~#1}, #2~(#3)}
\newcommand\ZP[3]{~Z. Phys. {\bf ~#1}, #2~(#3)}
\newcommand\NCBS[3]{~Nuovo Cimento B Serie {\bf ~#1}, #2~(#3)}
\newcommand\AAP[3]{~Astron. Astrophys.{\bf ~#1}, #2~(#3)}
\newcommand\MPLA[3]{~Mod. Phys. Lett. A.{\bf ~#1}, #2~(#3)}
\newcommand\PRIA[3]{~Proc. R. Irish Acad.{\bf ~#1}, #2~(#3)}
\newcommand\NCB[3]{~Nuovo Cimento B {\bf ~#1}, #2~(#3)}

\end{document}